\documentclass[amsmath, amssymb, aps, prx, reprint, onecolumn, longbibliography, floatfix, nofootinbib]{revtex4-2}
\usepackage[utf8]{inputenc}
\usepackage[urlcolor=blue, colorlinks=true, citecolor=blue]{hyperref}

\usepackage{amsmath}
\usepackage{amsmath,amssymb}
\usepackage{nicematrix}
\usepackage{gensymb}
\usepackage{tabularx}
\usepackage{algorithm}
\usepackage{physics}
\usepackage{algpseudocode}
\usepackage{multirow}
\usepackage{adjustbox}

\begin{document}
\title{Protein-protein docking using a tensor train black-box optimization method}

\author{Dmitry~Morozov}
\email{dimo@terraquantum.swiss}
\address{Terra Quantum AG, Kornhausstrasse 25, 9000 St. Gallen, Switzerland}

\author{Artem~Melnikov}
\address{Terra Quantum AG, Kornhausstrasse 25, 9000 St. Gallen, Switzerland}

\author{Vishal~Shete}
\address{Terra Quantum AG, Kornhausstrasse 25, 9000 St. Gallen, Switzerland}

\author{Michael~Perelshtein}
\email{mpe@terraquantum.swiss}
\address{Terra Quantum AG, Kornhausstrasse 25, 9000 St. Gallen, Switzerland}


\begin{abstract}
Black-box optimization methods play an important role in many fields of computational simulation.
In particular, such methods are often used in the design and modelling of biological systems, including proteins and their complexes with various ligands. This work is mainly focused on the protein-protein docking that plays a key role in modern drug-design workflows. We develop a black-box approach for such docking problems using a novel technique based on the tensor-train decomposition of high-dimensional interaction functions. Our method shows an advantage in terms of the discovered global minima and has a high potential for further implementation on a wide range of devices, including graphical processing units and quantum processing units.
\end{abstract}

\maketitle

\section{Introduction}

Protein-protein interactions play a crucial role in organisms \cite{eisenberg2000protein}. Understanding how and why individual molecules could form complexes of a very distinct structure is challenging. There are experimental methods to determine the structures of complexes, like crystallography and CryoEM techniques, which are typically deposited into the Protein Data Bank (PDB) \cite{berman2000protein}. However, they can cover only a very tiny fraction of all biological complexes. To overcome these limitations, computational methods for predicting and building protein complexes are becoming more popular, as a complementary addition to the experimental techniques.

Molecular docking is one of the major methods in studying protein-protein interactions at the atomistic scale \cite{meng2011molecular,de2016molecular}. 
Essentially, the method aims to sample the geometrical space of two or more proteins and search for the best poses of one protein (the ligand) with respect to the other protein (the receptor) using scoring functions. 
The main challenges are: (i) to design the scoring function to give an acceptable precision; and (ii) to efficiently sample the conformational space to find the best possible structures.

Today’s most widely used sampling methods are based on the Fast Fourier Transform (FFT). They are capable of generating geometrically complementary rigid-body docking poses \cite{pagadala2017software}.
However, they are typically limited in the use of very specific scoring functions and thus, in many cases require an \textit{a posteriori} recalculation of the scores for the obtained structures with additional external software \cite{cheng2007pydock}. On the other hand, there are existing black-box approaches for searching minimas in high-dimensional spaces. As an example, swarm intelligence algorithms, such as Particle Swarm Optimization (PSO) \cite{li2010detection} and Glowworm Swarm Optimization (GSO) methods \cite{krishnanand2009glowworm}, have been successfully applied to the protein docking problem using a large variety of scoring functions \cite{jimenez2018lightdock}.

In this work, we are aiming to apply another class of black-box global optimization algorithms based on the tensor train (TT) decomposition of the high-dimensional scoring (or energy) function \cite{TT_main_paper} that were developed over the last years \cite{sozykin2022ttopt}. 
The expected advantages of such methods is their agnosticism to the initial conditions and high-efficiency in sampling even very heterogeneous coordinate spaces that have a large number of local minima in their structure. These properties make such methods promising candidates for exploring interactions in biological complexes.

Another advantage of the tensor train methods developed by Terra Quantum is their capability for {\bf efficient implementation on both CPUs and GPUs}.
As can be seen from the results in this paper, usage of these algorithms already produces highly promising results. 
Moreover, the tensor train algorithms possess a structure similar to quantum circuits \cite{huggins2019towards, rudolph2022_shallow_decomposition}, which paves the way towards their future implementation on Quantum Processing Units (QPUs) once such QPUs are of sufficient scale and fidelity \cite{kordzanganeh2022benchmarking}. 
Thus, we are able to harness some advantages of our {\it Quantum Software} approach \cite{perelshtein2022practical}, which will grow once quantum hardware matures and numerical methods of quantum information processing via tensor networks advances.

\section{Methods}

Three different algorithms for solving docking problem were employed in the study: the glowworm swarm optimization (GSO) algorithm as implemented in the LightDock \cite{roel2020lightdock} package; the tensor train based minima search implemented in the ttpy library (TTPY) \cite{ttpy}; and our own tensor train based black-box optimization method packaged in a Python library (TTOPT), which will be described in the following section.

\subsection{Tensor Train assisted docking}

\begin{figure}[ht]
    \centering
    \includegraphics[width=0.9\linewidth]{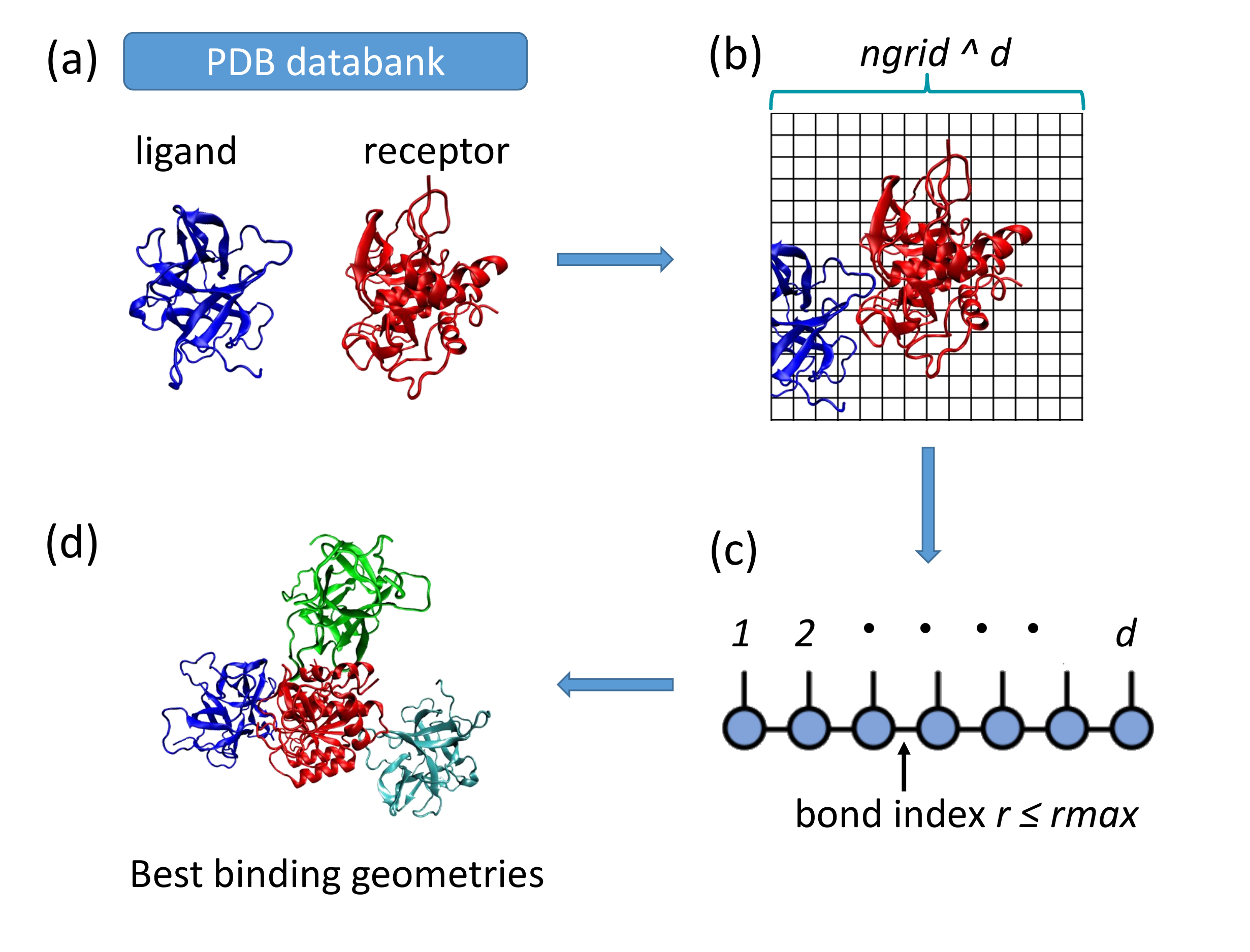}
    \caption{Workflow for performing docking simulations with the TTOPT approach. (a) - structure preparation, (b) - grid and target function definition, (c) -  decomposition of the target function into a tensor train and search for optima on a grid, (d) - clusterization of the obtained minima geometries.}
    \label{fig:docking_workflow}
\end{figure}

Here we will describe our workflow for performing docking simulations using the optimization algorithm based on the Tensor Train (TT) decomposition \cite{TT_main_paper, sozykin2022ttopt, zheltkov2020global}. The overall scheme is shown in Fig.~\ref{fig:docking_workflow}. Initially, as in the other docking protocols, the  preparation and cleanup of the ligand and receptor structures should be performed. At that stage, we also decide on the target (scoring) function used to evaluate the docking poses of the ligand molecule with respect to the receptor. This stage could also involve fixing the missing residues and side-chains, adding hydrogenes, building molecular topologies and 
partially relaxing the structures.

For the next stage, one would need to define a grid spanned around the receptor Fig.~\ref{fig:docking_workflow}b in which optimization of the target function with respect to the ligand position will be performed. It could be both minimization of the energy or maximization of the scoring. In general, our TT algorithm requires discretization of the search on a uniform grid. We define $d$ as the number of variables (dimensions) and $ngrid$ as the grid size in one dimension. In the current work, we use rigid-body models of the receptor and ligand proteins meaning that the total search space has 7 coordinates in total: 3 coordinates for the ligand's center of masses and 4 coordinates for its rotation quaternion. It should be noted that while here we are focused mainly on rigid-body protein-protein docking, the same approach is also applicable to other systems such as small molecules and nucleic acids, 
where the flexibility of these systems could be straightforwardly treated as additional dimensions in the TT decomposition.

The TTOPT algorithm would be used to search for the possible optima of the target function in the defined grid space. Unlike brute-force grid search, this algorithm does not estimate the cost function at all points of the grid but dynamically provides the next set of evaluating points in the search space based on the knowledge accumulated during all previous evaluations. The TTOPT algorithm 
applies the cross-approximation technique to the tensors \cite{tt_cross}, which in turn is based on the {\tt MaxVol} routine \cite{maxvol}. Essentially, it performs a decomposition of the multi-dimensional target function into a compressed form as the product of small tensors connected into a chain (Fig.~\ref{fig:docking_workflow}c) and expanded on a grid. 

In addition, the iterative power algorithm by Soley \textit{et.al.}\cite{soley2021iterative} is used for the global minima search within the TT representation. The total number of the target function evaluations required by our approach to find global minima could be approximated with $O(nswp*d*ngrid*rmax^2)$, where $d$ is the number of dimensions, $ngrid$ is the grid-size, $rmax$ is the maximum rank of the tensors with which the algorithm tries to approximate the target function via tensor train decomposition and $nswp$ is the number of iterations which the algorithm performs to improve the decomposition. Overall, our TTOPT approach scales much better with the system size than the $ngrid^d$ brute force grid search and shows much better parallelization capabilities than other black-box methods, such as Bayesian optimization or genetic algorithms. 
We implement the TTOPT core functional using C++ and package the whole framework in a Python library.

Finally, after the global optima search, we perform an analysis of the obtained receptor-ligand complex structures in terms of the target function values and clusterize them 
according to the root mean square deviation (RMSD) of the ligand C$\alpha$ protein backbone atoms (Fig.~\ref{fig:docking_workflow}d). Obtained in such a way, the best complexes from all found clusters could be further used to perform a comparison with the experimental data or for conducting simulations such as molecular dynamics, binding free-energy evaluations or direct mutagenesis using more elaborate approaches.

\subsection{Systems for docking}

To perform docking simulations, we have used 10 systems from the CAPRI scoring benchmark \cite{lensink2014score_set}. The Protein Data Bank (PDB) structures along with the chain identifiers, which have been used as the receptor and ligand parts of the respecting complexes to perform the simulations, are summarized in Table~\ref{tab:systems}. The fast implementation of the DFIRE energy function \cite{zhang2004accurate} provided in the LightDock software package was used as a target function for performing global minima search. 

\begin{table}[ht!]
\caption{Systems considered in this work. Chain - ID taken from the original PDB, Length - number of the residues in the Chain, Weight - molecular weight of the Chain in kDa }
\begin{tabular}{ccllcll}
\hline
\multirow{2}{*}{PDBID} & \multicolumn{3}{c}{Receptor}               & \multicolumn{3}{c}{Ligand}                \\ \cline{2-7} 
                       & \multicolumn{1}{l}{Chain} & Length & Weight & \multicolumn{1}{l}{Chain} & Length & Weight \\ \hline
3BX1                   & A                         & 269    & 26.7  & C                         & 181    & 19.9 \\
3U43                   & B                         & 132    & 15.3  & A                         & 92     & 10.0 \\
3R2X                   & AB                        & 492    & 56.8 & C                         & 82     & 10.6 \\
3Q87                   & B                         & 164    & 19.2  & A                         & 122    & 14.2 \\
2WPT                   & B                         & 113    & 15.2  & A                         & 82     & 9.9 \\
3E8L                   & A                         & 223    & 23.3  & C                         & 176    & 20.2 \\
3FM8                   & C                         & 364    & 45.6  & B                         & 95     & 14.0 \\
2W83                   & A                         & 161    & 19.0  & CD                        & 129    & 17.7 \\
2REX                   & B                         & 171    & 21.9  & A                         & 107    & 13.3 \\
2VDU                   & D                         & 376    & 51.3  & F                         & 201    & 29.7 \\ \hline
\end{tabular}
\label{tab:systems}
\end{table}

In the GSO optimization, 128 swarms have been simulated. Each swarm consisted of the 600 glowworms randomly placed at the beginning of the simulation. 200 algorithm iterations have been performed for each swarm. In addition, the best glowworm in each swarm has been minimized for 100 steps using the Powell minimization method \cite{powell1964efficient} at each GSO iteration. That resulted in $\sim140000$ (i.e., $600*200 + 200*100)$ total evaluations of the target function.

The tensor train based methods used the same parameters for the generation of the target function approximation with: the maximum rank of the tensor-train expansion $rmax=3$; the number of grid points $ngrid=64$ along each of the 7 dimensions (3 for the center of masses translation and 4 for the rotation quaternion); and the total number of iterations $nswp=10$. The parameters were chosen such that the methods resulted in $\sim120000$ target function evaluations. To be consistent with the GSO algorithm, we performed the TT-based minima search 128 times as well, starting with different random number seeds. In addition, for the TTPY method, we performed local energy minimization of the final structures for 100 steps using the same Powell approach. In TTOPT, we performed additional local minimizations for 20 steps for the best found points at the end of each iteration.

\begin{adjustbox}{center, caption={Scores evaluated with the DFIRE energy function for the five top clusters of structures obtained for docking complexes with Terra Quantum's TTOPT (blue) and the TTPY (orange) and GSO (gray) approaches. Higher scores indicate more optimal solutions.}, label={fig:docking_results}, nofloat=figure, vspace=\bigskipamount}
    \includegraphics[width=15cm]
    {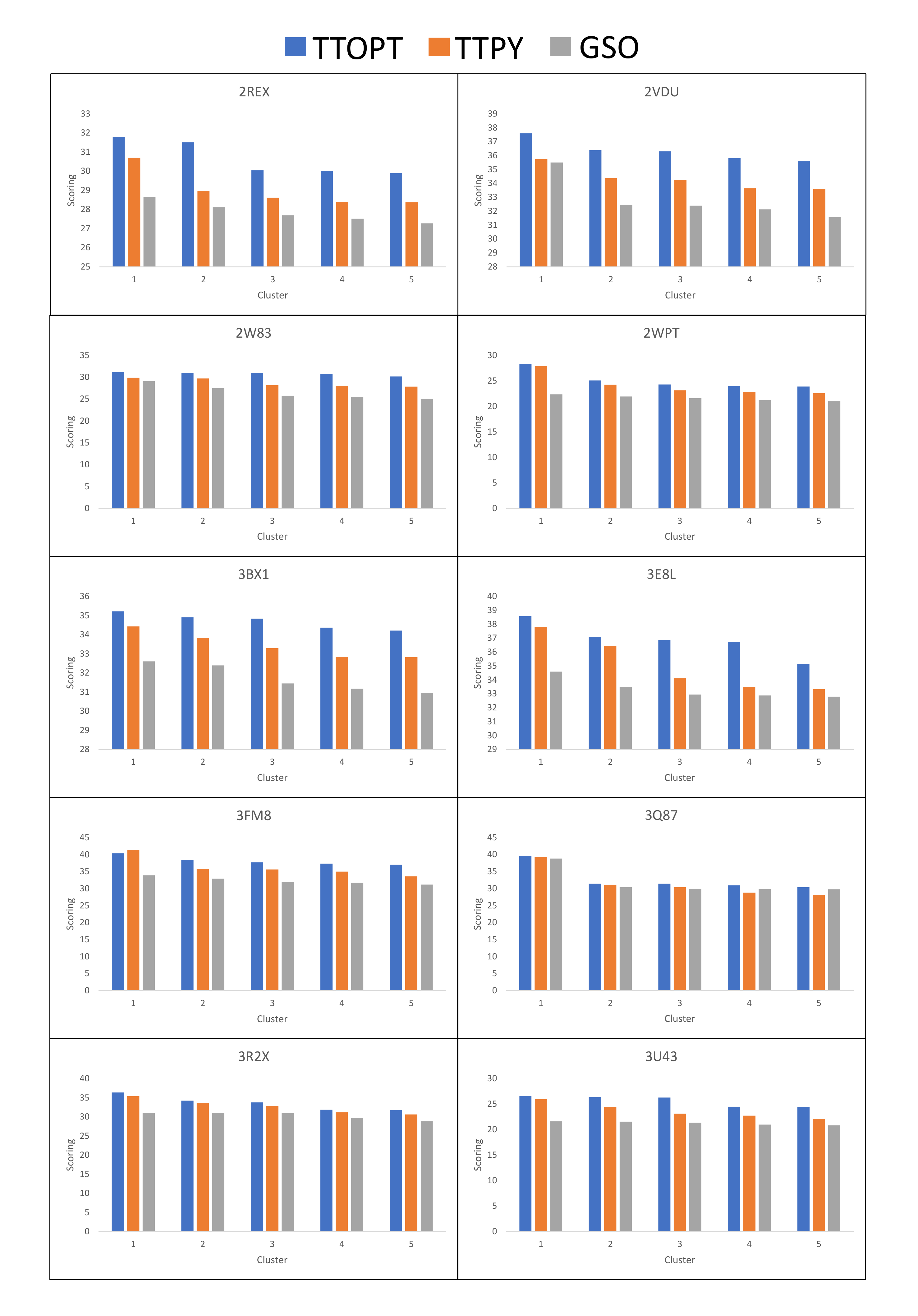}
\end{adjustbox}

All simulations were performed in the modified version of the LightDock package that was changed in order to enable usage of the all three global minima search methods. 

\section{Results and Discussion}
\label{sec:ResultsAndDiscussion}

\begin{figure}[ht]
    \centering
    \includegraphics[width=1\linewidth]{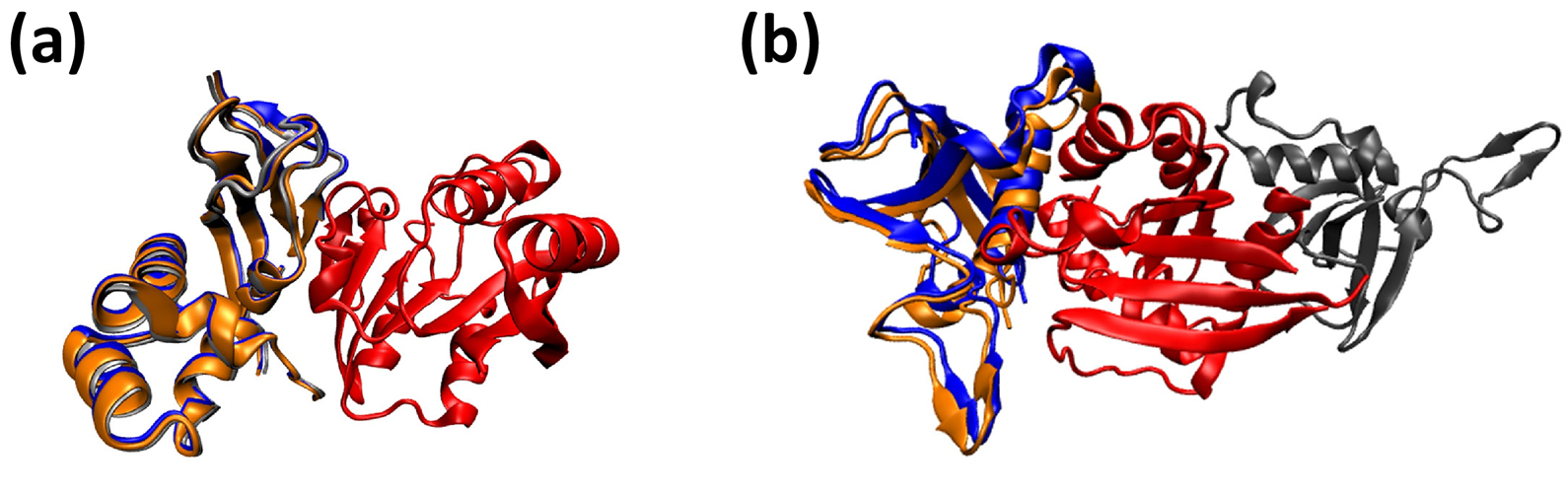}
    \caption{Overlap of the best structures  for the docking ligand protein found by employing the GSO (gray), TTOPT (blue) and TTPY (orange) algorithms. The receptor protein is shown in red. For the 3Q87 panel (a), the complexes are nearly identical. For the 2REX panel (b), the tensor-train based approaches found better minima. 
    }
    \label{fig:docking_pose_comp}
\end{figure}

The simulations were carried out using 32 CPU cores on the Google Cloud Platform. 
Afterwards, the simulations results were post-processed and clusterized. For tensor-train based methods, the positions of the C$\alpha$ protein backbone atoms for the best ligand positions found in each of the 128 simulations were generated and clusterized using RMSD criteria of 4\AA. For the GSO approach, the five best found structures of the receptor-ligand complex for each swarm were generated and further clusterized using the same 4\AA~RMSD criteria for the C$\alpha$ atoms. The five best-scoring clusters were used for comparisons of each complex. The results are shown in Fig.~\ref{fig:docking_results}. In the majority of cases, TTOPT outperforms GSO quite significantly in terms of the found global minima. It also outperforms the TTPY minimizer for most cases.
However, in complex 3Q87, we observed almost the same results for all three methods. Inspection of the found geometries for the best-scoring complexes shows that they are nearly identical as presented for the top cluster on Fig.~\ref{fig:docking_pose_comp}a. But in the other studied systems, the tensor-train based approaches tend to find better minima than GSO as shown for the 2REX system in Fig.~\ref{fig:docking_pose_comp}b.\\

\section{Conclusion}

We have presented a novel protein–protein docking approach based on the tensor-train decomposition of multi-dimensional scoring function landscapes. Our own TTOPT algorithm was implemented in the LightDock package. It demonstrated promising capabilities for sampling the translational and rotational space of rigid body protein–protein docking simulations. We demonstrated that our method outperforms LightDock's native GSO algorithm as well as the tensor-train based black-box global optimizer implemented in the ttpy library. 
Due to the nature of our optimization approach, it could also be employed for other chemical and biological modeling problems, such as the folding of proteins and nucleic acids. More broadly, it could be used in machine learning pipelines and for the optimization of hyperparameters \cite{sagingalieva2022hyperparameter}, for instance in the prediction of drug responses \cite{sagingalieva2022hybrid}. 

\section{Acknowledgments}
We are grateful to Prof. V. Vinokur and Dr. Y. Alexeev from Argonne National Laboratory for valuable discussions and suggestions. 

\clearpage

\bibliography{lib}
\bibliographystyle{unsrt}

\end{document}